\documentclass[conference]{IEEEtran}

\usepackage{color, bold-extra, pifont, paralist, calc, amsmath, mathtools, xspace, graphicx, url}
\usepackage{flushend}
\usepackage{xcolor}
\usepackage{soul}

\newcommand{\subpar}[1]{\textbf{\textit{#1. }}}
\graphicspath{{./img/}}
\newcommand{\cmark}{\ding{51}}
\newcommand{\xmark}{\ding{55}}

\definecolor{mod_colour}{HTML}{990000}
\newcommand{\rew}[1]{#1}
\newcommand{\rw}[1]{#1}
\newcommand{\switch}{{entry switch}\xspace}

\begin{document}
\title{Know Your Enemy: \\Stealth Configuration-Information Gathering in SDN}

\author{
\IEEEauthorblockN{Mauro Conti}
\IEEEauthorblockA{University of Padova\\
conti@math.unipd.it}
\and
\IEEEauthorblockN{Fabio De Gaspari}
\IEEEauthorblockA{Sapienza University\\
degaspari@di.uniroma1.it}
\and
\IEEEauthorblockN{Luigi V. Mancini}
\IEEEauthorblockA{Sapienza University\\
mancini@di.uniroma1.it}
}

\IEEEoverridecommandlockouts
\makeatletter\def\@IEEEpubidpullup{9\baselineskip}\makeatother
\IEEEpubid{\parbox{\columnwidth}{Permission to freely reproduce all or part
    of this paper for noncommercial purposes is granted provided that
    copies bear this notice and the full citation on the first
    page. Reproduction for commercial purposes is strictly prohibited
    without the prior written consent of the authors 
    (for reproduction of an entire paper only), and
    the author's employer if the paper was prepared within the scope
    of employment.  \\
}
\hspace{\columnsep}\makebox[\columnwidth]{}}

\maketitle

\begin{abstract}
Software Defined Networking (SDN) is a network architecture that aims at providing high flexibility through the separation of the network logic from the forwarding functions. 
The industry has already widely adopted SDN and researchers thoroughly analyzed its vulnerabilities, proposing solutions to improve its security. 
However, we believe important security aspects of SDN are still left uninvestigated. 

In this paper, we raise the concern of the possibility for an attacker to obtain knowledge about an SDN network. %
In particular, we introduce a novel attack, named \emph{Know Your Enemy} (KYE), by means of which an attacker can gather vital information about the configuration of the network.
\rew{This information ranges from the configuration of security tools, such as attack detection thresholds for network scanning, to general network policies like QoS and network virtualization. 
Additionally, we show that an attacker can perform a KYE attack in a stealthy fashion, i.e., without the risk of being detected}. 
We underline that the vulnerability exploited by the KYE attack is proper of SDN and is not present in legacy networks. 
    To address the KYE attack, we also propose an \rew{active defense} countermeasure based on network flows obfuscation, which considerably increases the complexity for a successful attack. 
Our solution offers provable security guarantees that can be tailored to the needs of the specific network under consideration.
\end{abstract}

\section{Introduction}
Software Defined Networking (SDN) is a network architecture proposed in recent years to address the shortcomings of traditional architectures. 
SDN posits that the implementation of network functions and the control logic of the network are two separate concepts, and should therefore be separated in different entities. 
To this end, SDN introduces the concepts of \emph{data plane} and \emph{control plane}: 
the data plane is comprised of the physical network devices (from here on, called switches) and implements the forwarding functionalities of the network, while the control plane manages the network logic and decision making process. 
In SDN, the control plane takes the decisions on how traffic flows are managed and pushes these decisions to the data plane, that will in turn enforce them. 
This separation between logical control and physical implementation of the network functions provides a high degree of flexibility, which is one of the main reason for the widespread adoption of SDN even amongst big companies~\cite{microsoft,Jain:2013:BEG:2534169.2486019}.

\rew{While the programmability of SDN allows for fast prototyping and high adaptability to different scenarios, it also opens new venues for attacks~\cite{Ambrosin:2015:LEM:2714576.2714612}. 
Indeed, while the decision making process is centralized in the control plane, the enforcement of the decision is distributed throughout all switches, which follow the rules pushed by the control plane. 
\rw{We show that, by exploiting this distributed policy-enforcement mechanism, an attacker can gather intelligence about the control logic of the network in a stealthy fashion.} 
In particular, \rw{to run the attack we propose in this paper, the adversary needs to have only a flow table side-channel, i.e., a way of learning which rules are installed, for a single switch (which we call \emph{\switch}, see Section~\ref{threat_model})}.
By analyzing the conditions under which a rule is pushed, and the type of such rule, an attacker can infer sensitive information regarding the configuration of the network. %
\rw{The final result is that through a single switch,} the attacker can gather information which, in a classical network, would have required \rw{access to numerous distinct devices, such as firewalls, intrusion detection/prevention systems, etc.} 
The information gathered can subsequently be exploited to mount different attacks, \rw{tailored to the target network, without being detected.}} 

To summarize, our contribution in this paper is as follows:
\begin{compactitem}
    \item We propose a novel, attack, the \emph{Know Your Enemy} (KYE) attack, that allows stealth intelligence gathering about the configuration of a target SDN network. 
        \rw{The information that an adversary can obtain ranges from configuration of security tools, such as attack detection thresholds for network scanning, to general network policies like QoS and network virtualization.} 
    \item We prove the feasibility and efficacy of the KYE attack through \rw{its implementation and} a thorough experimental evaluation on a test network. 
	\item We \rew{propose} a countermeasure to the KYE attack, based on obfuscation of inbound network flows. 
\end{compactitem}

\section{Preliminaries: OpenFlow}
\rew{SDN is a general architectural principle: it broadly defines general guidelines and overall architecture. 
When discussing SDN in real world scenarios, we refer to specific SDN implementations. 
In the remainder of this paper we focus on the OpenFlow implementation of SDN, due to its wide adoption in academia~\cite{7150550} and industry (including companies like Google and Microsoft~\cite{microsoft,Jain:2013:BEG:2534169.2486019}). 
However, it is worth noting that all our considerations are not specifically tied to OpenFlow, but hold true in general for SDN. }

\begin{figure}[ht]
	    \centering
	    \includegraphics[width=0.7\columnwidth]{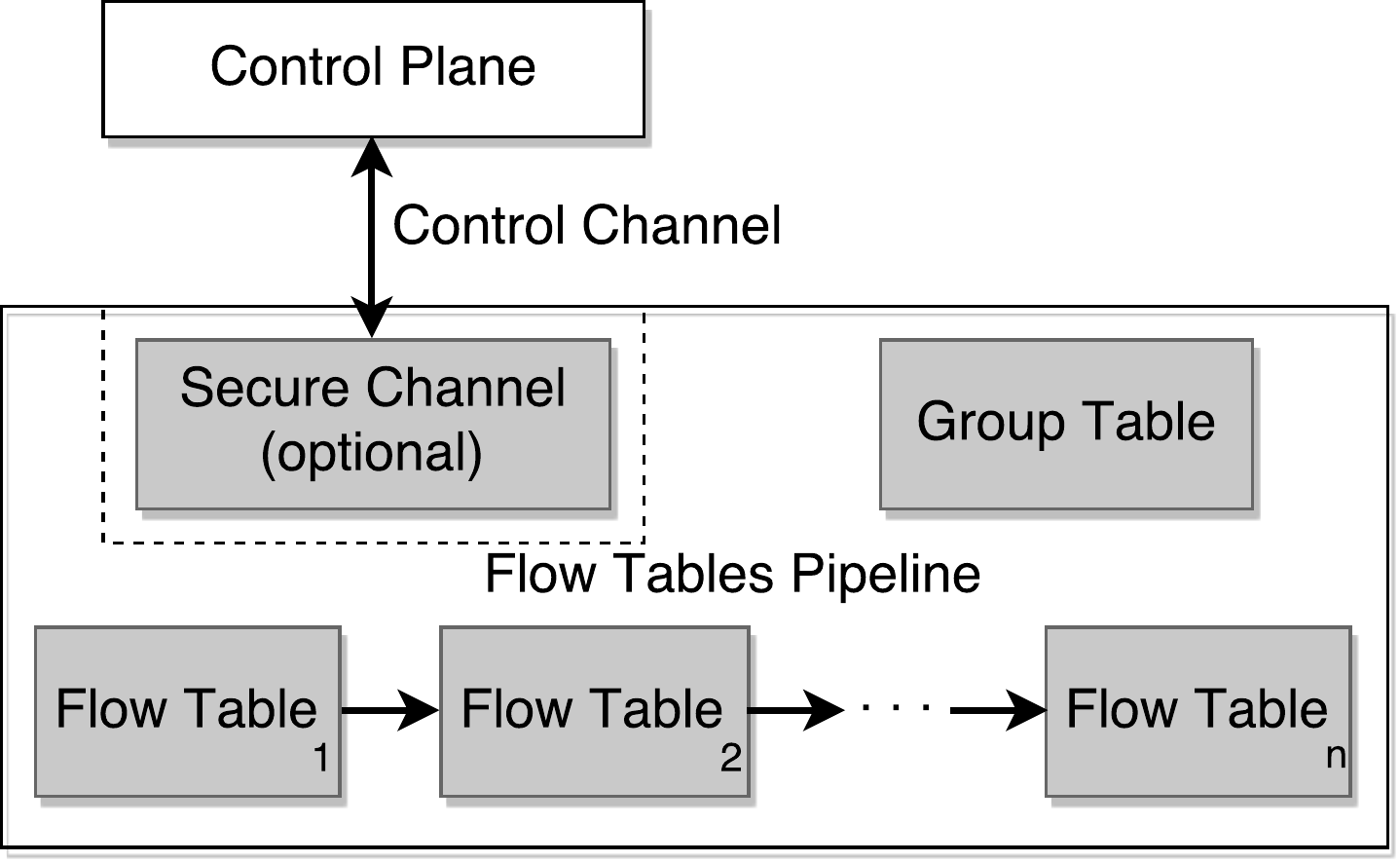}
	    \caption{Overview of the architecture of an OpenFlow switch.}
	    \label{fig:openflow}
	\end{figure}

The OpenFlow specification defines the communication protocol between the network controllers in the control plane and the network devices, called OpenFlow switches, in the data plane. 
Additionally, OpenFlow defines the general architecture and the functionalities of OpenFlow switches. 
Figure~\ref{fig:openflow} provides and overview of the architecture of an OpenFlow switch. 
In OpenFlow, each switch maintains a set of flow tables. 
Each of these flow tables contains a set of flow rules; flow rules are installed on OpenFlow switches by the control plane, and define what actions a switch needs to apply to a certain network flow. 
A flow rule can match one or more network flows, through the use of wildcards, and a network flow can be matched by one or more flow rules. 
Through combination of different rules and actions, the controller can instruct the switches to perform arbitrarily complex operations on the packets belonging to a given network flow. 
One of the main strengths of OpenFlow is the on-demand management of network flows: 
when a switch receives an inbound network flow for which it has no matching flow rule, it will contact the controller to request the installation of a new rule. 
This dynamic management of rules allows the controller to evaluate the current state of the network before deciding how to handle a network flow, which allows for the implementation of complex and flexible network policies. 
The communication between OpenFlow switches and the control plane can be either in plain text, or secured through TLS. 
\rw{The specification does not mandate TLS support, and in many cases commercial OpenFlow switches do not support it~\cite{Benton:2013:OVA:2491185.2491222}.} 

\section{Assumptions and SDN Issues}
\label{threat_model}
While having a logically centralized point of control allows to improve the decision-making process, distributing the policy enforcement introduces new problems with regard to information disclosure. 
Where network functions in legacy networks are relegated to the specific devices implementing them, providing higher control over access to their configuration, in SDN they are distributed throughout the OpenFlow switches. 
Indeed, network policies and functionalities like intrusion detection/prevention systems (IDS/IPS), network virtualization, or access control, are often enforced by the OpenFlow switches, through the application of the flow rules installed by the controller~\cite{6419708,sphinx,6362137}. 
Unfortunately, this behaviour considerably broadens the attack surface for an attacker. 

Indeed, as we show in this paper, by having a flow table side-channel on a single OpenFlow switch \rew(called \emph{\switch}), an attacker can gather a relevant amount of information regarding the behaviour and configuration of the SDN network the switch belongs to. 
\rew{A flow table side-channel is defined as any mean by which an attacker can learn or infer the content of the flow table of a switch. 
The KYE attack is independent from how this side-channel is obtained, and the detailed description of \rw{how to obtain it} is out of the scope of this paper. 
However, for completeness, we list here some of the  possibilities an attacker could leverage:
\begin{compactitem}
    \item Connect to a passive listening port on the \switch to retrieve the flow table.
        In fact, most OpenFlow switches can be remotely debugged by means of a passive listening port, which also allows retrieval of the flow table~\cite{Benton:2013:OVA:2491185.2491222,gregory14}. 
        For instance, an attacker could use the \texttt{dpctl} utility~\cite{dpctl} on the unprotected listener port of an HP Procurve~\cite{procurve}.
    \item Use Round Trip Time (RTT) variance to infer information about the content of the flow table~\cite{7480416,6733671}. 
    \item Sniff the control traffic. 
        Indeed, the use of TLS on the control channel is optional~\cite{openflow}, and many commercial switches do not support it~\cite{Benton:2013:OVA:2491185.2491222}. 
        Additionally, in most cases OpenFlow switches that support TLS do not implement \rw{certificate} authentication~\cite{gregory14}.
    \item \rw{Exploit a backdoor in the OS of the switch to read the flow table or decrypt the control traffic~\cite{juniper_backdoor}.}
    \item Use an hardware device to physically read the flow table~\cite{Carrier:2004:HMA:2307226.2307336}. 
        Such devices can acquire the content of the memory of the switch and copy it to an external destination. 
\end{compactitem}
It is worth noting that the attacker uses the flow table side-channel only to read the state of the flow table, therefore the overall state of the \switch is not modified in any way. 
This means that, even in case of a controller monitoring the integrity of the entire state of the switches, e.g., through direct query~\cite{Kamisinski:2015:FDM:2809826.2809833} or checksum~\cite{Benton:2013:OVA:2491185.2491222}, the flow table side-channel would not be detected.} 
Through \rw{the KYE attack} the attacker can learn all sorts of relevant information, from routing policies, to more complex and important behaviours regarding attack detection and defense mechanisms. 

Finally we would like to point out that the KYE attack exploits a structural vulnerability of SDN, which derives from the on-demand management of network flows, that in turn is one of the main features and strengths of this new network paradigm. 
Therefore, it is not something that can be easily modified or circumvented.

\subpar{Threat Model}
Our threat model assumes that the attacker has a flow table side-channel for a single OpenFlow switch. 
\rw{In particular, the only abilities of the attacker are (\emph{i}) sending packets through the target network and (\emph{ii}) using the side-channel to learn the flow rules that are installed on the \switch.}
\rw{The attacker can be either internal or external, does not have control over the switch, and does not modify the behaviour of the switch.} 
\rew{Furthermore, we assume that the switch only provides the functionality described in the OpenFlow standard~\cite{openflow}. 
Therefore, the attacker does not modify the software or add capabilities to the switch in any way.
}

\section{The KYE Attack}
\label{kye_overview}
\rew{From a high level view, the KYE attack follows a general attack strategy to obtain information on a target network. 
The details of the attack vary based on the specific information the attacker wants to gather (see Section~\ref{applied_kye}).
However, all instance of the KYE attack share a common kernel, as detailed below.} 
The main idea behind the attack comes from the observation that, in an SDN network, rules are pushed from the controller to the switches only when needed~\cite{openflow}. 
This behaviour holds true for all types of rules; for instance, the controller will push a flow rule to counter a denial of service (DoS) attack only when one is under way and detected. 
\rew{The KYE attack exploits this on-demand installation of flow rules, allowing an attacker to gather knowledge about which conditions trigger the installation of a given flow rule, as illustrated in Figure~\ref{fig:kye_attack}. 
The KYE attack is structured in two phases: (\emph{A}) the \emph{probing phase} and (\emph{B}) the \emph{inference phase}. 

In the probing phase (\emph{A}), which is repeated numerous times, the attacker attempts to trigger the installation of flow rules on the \switch (steps 1 through 5 in, Figure~\ref{fig:kye_attack}):
\begin{itemize}
    \item \sloppy{The attacker sends carefully crafted \emph{probing traffic} through the \switch in order to trigger the installation of new flow rules.} 
        The specific characteristics of the probing traffic depend on what kind of information the attacker is interested in learning (see Section~\ref{applied_kye} and Section~\ref{eval}). 
    \item Through the flow table side-channel, the attacker obtains the flow rule (if any) installed in response to the probing traffic.
\end{itemize}

In the inference phase (\emph{B}), the attacker analyzes the correlation between the probing traffic generated during the probing phase and the corresponding flow rules installed (step 6 in Figure~\ref{fig:kye_attack}). 
From this analysis, he can infer what network policy is enforced for specific types of network flows. 
For instance, if in response to scanning traffic generated in the probing phase the controller installs \emph{drop} rules, the attacker infers that the defense policy against network scanning is traffic filtering.
Additionally, by studying the features of the probing traffic, the attacker can potentially infer the trigger conditions for network policies that require a specific trigger before being activated (see Section~\ref{eval}).

	\begin{figure}[ht]
	    \centering
	    \includegraphics[width=\columnwidth]{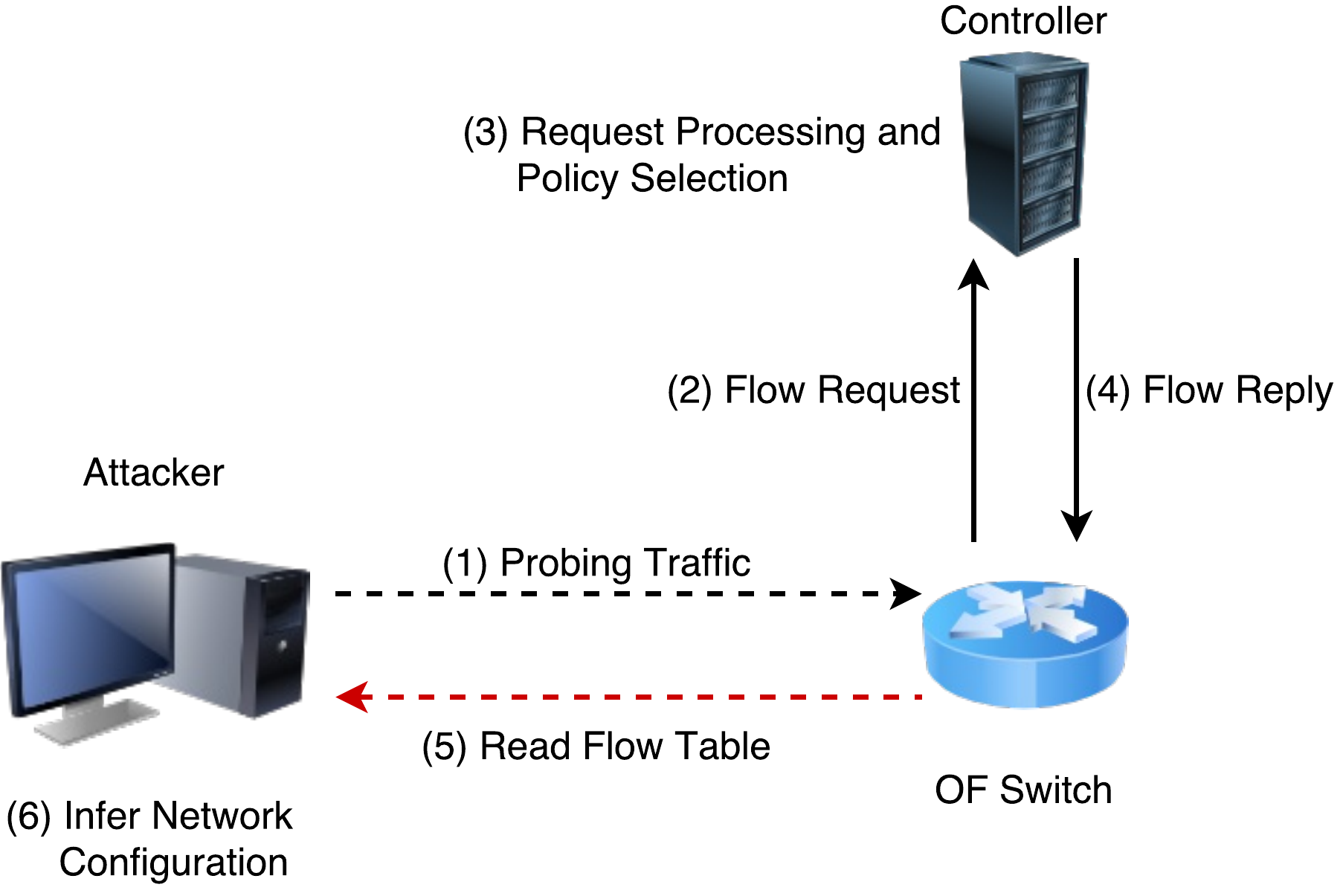}
	    \caption{Overview of a general KYE attack.}
	    \label{fig:kye_attack}
	\end{figure}

The final result is that, through the KYE attack, the attacker is able to learn the control logic of the SDN network regarding a specific type of network flow. 
Moreover, when the attacker is able to learn the trigger condition for a given network policy (e.g., \rw{policy against network scanning attacks}), he can then exploit this knowledge to perform additional attacks without triggering detection (see Section~\ref{exp_ac}). }
It is worth noting that the probing traffic might be directed to an end host that is not part of the target network. Indeed, the KYE attack only requires that the probing traffic is routed through a switch of the target SDN. 
This might be the case, for instance, of an ISP providing defense measures against network attacks to their clients. 
An attacker could perform a KYE attack to gather intelligence on the ISP network, using as a destination host for the probing traffic the hosts of \rew{one of the clients of the ISP}. 
Moreover, introducing probing traffic in the network does not expose the KYE attack to detection. 
Indeed, one of the strengths of the KYE attack is that it hides behind other attacks: 
for instance, an attacker performing a KYE attack to gather information related to DoS countermeasures will generate DoS traffic through the target network. 
When this traffic is detected, the network will simply conclude that a DoS attack is underway, and the KYE attack will remain undetected. 
Effectively, the probing traffic creates a smokescreen that hides the KYE attack to the network, making it extremely hard to detect in any reliable fashion. 

\section{KYE Instances}
\label{applied_kye}
\rw{As we discussed before, the KYE attack is a general attack strategy, and the details of the attack vary based on the specific information the attacker wants to gather.} 
In this section, we \rew{discuss different instances of the KYE attack, with respect to what type of information the attacker wants to obtain.} 
Section~\ref{kye_sec} presents the KYE attack in relation to the disclosure of security-related network configuration information. 
In Section~\ref{kye_sdn} we discuss how the KYE attack can be used to gather intelligence on SDN-related network configuration.
Finally, in Section~\ref{kye_gen} we analyze the use of the KYE attack to disclose general network configuration information. 
Additionally, we provide a non-exhaustive list of concrete examples of KYE attack, showing how an attacker can exploit the flow table side-channel to infer different configuration features of the network. 

\subsection{\sloppy Gathering Network Security Configuration Information}
\label{kye_sec}
When planning an attack, knowing what detection and defense mechanisms are used by the target network is obviously invaluable to an attacker. 
In this section we discuss how, through repeated probing and analysis of the flow table, an attacker can infer detection mechanisms and defense measures in place in an SDN network for different types of attack. 
\rw{Due to space limitations,} in our analysis we focus on some popular SDN-based defense mechanisms proposed in the literature~\cite{7226783,7150550}. 
\newline
\subsubsection{Worm Infection/Scanning}
\label{scanning_inference}
Scanning is one of the main preliminary intelligence gathering techniques, used by attackers to gather information about a given target network and by worms to detect vulnerable targets to spread the infection to. 
Through scanning, an attacker can learn about the number, type and address of hosts in a network, along with what services are offered on which port. 
This information is a prerequisite for mounting more complex attacks. 
Therefore, being able to detect and mitigate scanning is extremely important for any network. 

\subpar{KYE Attack}
In SDN, an attacker can infer information regarding the type of defense mechanisms used to mitigate scanning and, depending on the detection mechanism employed,  the detection threshold for scans. 
In order to infer such information, an attacker simply needs to send scanning probes from a spoofed address $IP_A$, varying the characteristics of the scan (e.g., increasing scanning rate, different duration of the scan, various success/failed connection ratios).
After each probe, the attacker reads the content of the flow table of the OpenFlow switch and takes note of the new flow rule installed in response to the probe. 
As long as the scanning is not detected, the flow rules installed simply instruct the switch to forward the traffic coming from $IP_A$ towards different exit ports, based on the destination address. 
When the scanning rate becomes high enough or the scanning activity lasted long enough for the attack to be detected~\cite{mehdi2011}, \rew{the controller will install flow rules implementing an appropriate defense measure against scanning.} 

\subpar{Detection and Defense Mechanism Inference}
Depending on the defense measure used by the network, different types of flow rules will be installed, for instance: traffic filtering~\cite{entropy_based}, rate limiting~\cite{mehdi2011,Twycross:2003:ITV:1251353.1251373}, honeypot redirection~\cite{FRESCO}, or whitehole network approaches~\cite{Ambrosin:2015:LEM:2714576.2714612,Shin:2013:ASV:2508859.2516684}. 
All these defense mechanisms require the installation of very specific flow rules, which differ from the normal flow rules installed when no attack is detected (see Section~\ref{defence_inference}). 
\rew{Consequently, by recognizing this change in the type of rules installed, the attacker can learn that the network scanning was detected.} 
Moreover, the flow rules required by these mechanisms are easily identifiable and, just by looking at what rule is installed, the attacker can infer the defense mechanism used by the network (see Section~\ref{defence_inference}). 
Finally, for some defense mechanisms that are activated on-demand by the controller, the attacker can also infer \rew{the traffic features that trigger detection by the network.} 
For instance, if we consider TRW-CB~\cite{Schechter2004}, which is one of the most frequently used anomaly detection algorithms~\cite{Ashfaq2008} and is  already implemented in SDN~\cite{mehdi2011,FRESCO}, the attacker can learn the ratio between successful/unsuccessful connections used as detection criteria through repeated probing. 
Once the attacker discovers the detection threshold, he is then able to carry out the network scan undetected in a second phase, by alternating unsuccessful scanning with successful connections. 

\rew{\subpar{Implementation}
In Section~\ref{exp_trw} we \rew{report on the implementation of this instance of the KYE attack.} 
We show that, for realistic detection and defense mechanisms, an attacker is able to learn both the scanning traffic features that trigger detection and the defense mechanism applied in response. }
\newline
\subsubsection{Denial of Service}
DDoS attacks are one of the most widespread type of attacks and their diffusion and sophistication increases every year~\cite{ddos_report}. 
Indeed, attackers often use DDoS attacks as a smokescreen to cover data theft and malware installation on target systems~\cite{ddos_report}.
Moreover, the financial damage these attacks cause can range in hundreds of thousands of dollars in peak hours~\cite{ddos_report2}. 
As a consequence, most organizations adopt one or more DDoS detection and mitigation system~\cite{ddos_report2}. 
In the context of SDN, DDoS detection schemes tend to employ lightweight mechanisms, such as threshold and entropy-based systems~\cite{entropy_based,sphinx} to avoid overloading the controller. 
Other more complex and computationally expensive approaches do exist, like~\cite{5735752}, where the authors employ machine learning techniques to detect possible DoS attacks.
Depending on which detection mechanism is used, an attacker can perform a KYE attack to learn if a DoS detection mechanism exists, what defense measure is applied by the network and potentially even the detection criteria employed.

\subpar{KYE Attack}
The KYE Attack for DoS detection is very similar to the one used for network scanning; 
in the probing phase, the attacker starts a DoS with a low attack rate, simulating a behaviour that is as close as possible to that of a legit client, then gradually increases the profile of the attack.
Throughout the attack, the attacker monitors the flow rules installed by the controller on the OpenFlow switch, looking for a change in the flow rules pushed. 
Indeed, under normal circumstances the controller will simply instruct the switch to route the traffic towards the destination host. 
However, when the DoS attack is detected, the controller pushes different rules based on the defense mechanism employed by the network, allowing the attacker to learn that the DoS was detected. 

\subpar{Detection and Defense Mechanisms Inference}
Defense mechanisms proposed against DoS attacks, like traffic redirection~\cite{Mahimkar:2007:DTN:1973430.1973454}, rate limiting~\cite{Twycross:2003:ITV:1251353.1251373,mehdi2011} or traffic filtering~\cite{entropy_based}, require very specific flow rules that an attacker can easily distinguish from normal routing rules (see Section~\ref{defence_inference}). 
Therefore, as soon as these security flow rules are installed on the monitored OpenFlow switch, the attacker will know that the DoS was detected. 
By analyzing the specific flow rules installed, he can also infer the defense mechanism applied (see Section~\ref{defence_inference}).
Additionally, when the attack is detected, the attacker can try to infer the detection criteria used by the network. 
Indeed, for certain type of detection mechanisms such as threshold~\cite{sphinx} or entropy-based~\cite{entropy_based}, it is possible to find a good approximation of when exactly an attack is detected. 
In order to learn these detection thresholds, an attacker can repeat the probing phase several times varying the characteristics of the attack. 
Upon detection, the attacker will log such characteristics, like duration, attack rate and number of packets sent from each IP address, for instance. 
After obtaining a sufficiently large sample, the attacker can look for correlations between the characteristics of the detected attacks to learn approximately what values trigger the detection. 
Even in case of more complex detection systems based on machine learning~\cite{5735752}, the attacker can still obtain knowledge about the traffic features used for detection. %
Indeed, previous work on the area of machine learning shows that it is possible to infer meaningful information about the training set of a classifier~\cite{Ateniese:2015:HSM:2829869.2829870}, which in our case would reveal information about which flows are considered malicious or benign. 
\newline
\subsubsection{Access Control}
Access control mechanisms, like firewalls, are the first and most basic defense mechanism used by networks to enforce security policies~\cite{4623689}. 
Through its centralized view of the network and distributed enforcement of rules, SDN provides the optimal functionality to implement a consistent distributed firewall in the network~\cite{Hu:2014:FBR:2620728.2620749,6779061,Wang2013}.
Given that such access control systems are the first defense mechanism that an attacker needs to bypass before attacking a network, learning the exact configuration of such devices would provide a huge advantage in preparing an attack. 
While this would be extremely challenging in classical networks, due to the fact that access control rules are relegated only to specific security devices, this task is considerably simpler in SDN. 

\subpar{KYE Attack}
By performing a KYE attack, the attacker can infer all the access control policies he is interested in by simply probing the monitored OpenFlow switch. 
In its most basic form, the attacker will send a probe packet to test any given access control policy. 
For instance, the attacker can try to connect to a protected service using a set of different IPs, in order to understand which subnets are allowed to access that service. 
For each of these probes the controller will push either a forwarding flow rule, to forward the packets to their destination, or a \emph{drop} rule if the access is not allowed~\cite{4623689}. 
By repeating the probing for all interesting services and using different source IP addresses, the attacker can map which addresses (or address ranges) are allowed towards/from a certain critical service. 

\subpar{Defense Mechanism Inference}
For access control enforcement, \rew{the policy that is most used in general} is to drop unauthorized traffic flows~\cite{4623689}. 
If such defense mechanism is in place, the attacker is able to recognize it immediately just by reading the \rew{new flow rule installed} (see Section~\ref{defence_inference}). 
SDN also allows for more complex defense mechanisms to enforce access control, like traffic redirection towards a honeypot/IDS for instance~\cite{Mahimkar:2007:DTN:1973430.1973454,FRESCO}. 
As we discuss in Section~\ref{defence_inference}, the attacker can easily identify even this more complex mechanisms just through observation of the OpenFlow switch flow table. 

\rew{\subpar{Implementation}
In Section~\ref{exp_ac} we \rew{report on the implementation of this instance of the KYE attack.} 
We show that, given an access control mechanism, an attacker is able to learn the complete access control matrix enforced by the controller.}

\subsection{\sloppy Gathering SDN-Related Configuration Information}
\label{kye_sdn}
Through a KYE attack, an attacker can infer vital information about SDN-related network configuration, such as flow table management~\cite{Al-Fares:2010:HDF:1855711.1855730,Curtis:2011:DSF:2018436.2018466} and control plane scalability measures~\cite{Ambrosin:2015:LEM:2714576.2714612,Shin:2013:ASV:2508859.2516684}. 
Knowledge about the configuration of such critical systems provides an attacker with a wider attack surface, as well as allowing him to better focus his resources during an attack.
\newline
\subsubsection{Flow Table Saturation}
Flow rules, and therefore flow tables, are the core enablers of SDN in OpenFlow.
While fine-grained flow rules allow for targeted policy enforcement, the number of flow rules that can be installed on OpenFlow switches is limited. 
If the flow rule limit is reached (e.g., as a result of a deliberate saturation attack), the switch will not be able to accept rules for new inbound network flows, which will be ignored. 
To mitigate this vulnerability, the control plane can employ wildcard rules to aggregate the management of multiple network flows with a single flow rule~\cite{Al-Fares:2010:HDF:1855711.1855730,Curtis:2011:DSF:2018436.2018466}.
From the point of view of an attacker, learning under which conditions the controller uses wildcard rules and, more importantly, how to force it to install targeted flow rules, is essential to successfully mount saturation attacks. 

The main consideration behind aggregate network flow management is that, in general, it is important to route flows on an individual basis only under certain conditions. 
In particular, for the purpose of network engineering, this conditions are generally related to the weight of a specific network flow; network flows that surpass a certain threshold (be it a data rate~\cite{Al-Fares:2010:HDF:1855711.1855730} or size~\cite{Curtis:2011:DSF:2018436.2018466} threshold), are marked for individual routing through targeted flow rules.

\subpar{KYE Attack}
Through a KYE attack an attacker can infer the thresholds used to classify network flows, allowing him to flood the network with flows for which the controller will generate targeted flow rules. 
Indeed, similarly to a KYE attack against threshold based detection systems (see Section~\ref{scanning_inference}), an attacker can easily detect if such aggregation systems are in place and, if so, what thresholds they employ. 
As a first step, the attacker reads the OpenFlow switch flow table and creates a new network flow not matching any flow rule present. 
If network flow aggregation is in use, the new flow rule installed on the switch will be a wildcard flow rule, otherwise it will be a flow rule targeted specifically at the new network flow. 
If a wildcard rule is installed, the attacker can then increase the data rate transmission for that network flow until a new, flow-specific rule is installed by the controller. 
At this point, the attacker knows the exact attack rate and/or size of the flow (cumulative amount of data transmitted) required for the controller to install a targeted flow rule. 
\newline
\subsubsection{Control Plane Scalability}
An important issue in SDN lies within the scalability of its control plane.  
Indeed, previous research demonstrates that, due to the communication required between data and control plane, SDN is subject to a particular type of DoS attack known as control plane saturation attack~\cite{Ambrosin:2015:LEM:2714576.2714612,Shin:2013:ASV:2508859.2516684}. 
This attack aims at overloading the control plane with flow requests by flooding OpenFlow switches with a high number of unique network flows. 
To mitigate this problem, researchers proposed data plane extension modules aimed at validating the inbound traffic before forwarding a flow request to the controller~\cite{Ambrosin:2015:LEM:2714576.2714612,Shin:2013:ASV:2508859.2516684}. 
For an attacker, detecting if such countermeasures exist is fundamental in order to successfully attack the target network. 
The core idea behind these countermeasures is to perform the traffic validation at the data plane, before contacting the controller. 
In~\cite{Shin:2013:ASV:2508859.2516684}, for instance, the traffic validation is done through the use of SYN proxy as soon as an OpenFlow switch receives a SYN packet. 

\subpar{KYE Attack}
An attacker can detect the presence of this kind of defense mechanisms by means of the KYE attack; 
indeed when SYN proxy techniques are used, the OpenFlow switch answers to any SYN packet without a flow rule being installed by the controller. 
An attacker can exploit this behaviour by generating a new connection request (SYN packet) for which there are no flow rules already installed in the flow table. 
Upon receiving a SYN-ACK response, if there are still no matching flow rules for the generated connection request, then SYN proxy countermeasures are in place in the OpenFlow switches. 
This behaviour exposes the use of SYN proxy techniques at the data plane level to the attacker, who is then free to attack the network through vulnerabilities of the SYN proxy approach~\cite{Ambrosin:2015:LEM:2714576.2714612}.

\subsection{\sloppy Gathering General Network Configuration Information}
\label{kye_gen}
Beside security and SDN-related applications and mechanisms, networks have \rw{numerous} general functionalities and policies ranging from network virtualization~\cite{6362137,Sherwood:2010:PNT:1924943.1924969} to traffic shaping and quality of service (QoS)~\cite{6411795,6116083}. 
In this section we discuss how an attacker can gather knowledge about such functionalities through the KYE attack. 
\newline
\subsubsection{Network Virtualization}
Network virtualization is a key component to enable sharing of physical resources. 
In this context, the abstraction and programmability offered by SDN provide great advantages in the implementation of such network virtualization techniques. 
Indeed, there have been several proposals in the literature on how to employ SDN to obtain efficient resource sharing in the context of network virtualization~\cite{6362137,Sherwood:2010:PNT:1924943.1924969}. 
Although they provide considerable benefits, virtualization and resource sharing also create new problems and vulnerabilities. 
In fact, sharing physical hardware between multiple tenants can lead to leakage of critical information~\cite{Ristenpart:2009:HYG:1653662.1653687}. 
From the point of view of an attacker, learning if some type of network virtualization is applied and, if so, gathering as much information as possible from it, is therefore clearly important. 
When we consider SDN-related virtualization schemes like FlowN~\cite{6362137}, network virtualization is implemented through flow tagging: edge routers tag the traffic with an additional header, which is used by the control plane to map a flow with the routing logic specified by the appropriate tenant. 

\subpar{KYE Attack}
Through a KYE attack an attacker can learn, for instance, how network flows are managed by other tenants in the network or which tenants he is sharing a physical machine with. 
Learning flow management for other tenants is straightforward, since all the flow rules for all tenants are installed in the same set of flow tables. 
The attacker can just read the various flow rules present in the flow table of the OpenFlow switch, differentiating them based on the header for the specific flow (which indicates the tenant owning that flow).  

In order to gather knowledge about which tenants are co-resident on the same physical machine, an attacker just needs to generate probe packets directed to a service run by the tenants he is interested in.
If the target tenant is co-resident on the same physical machine as the attacker, the new flow rule installed in response to the probe will instruct the switch to output the traffic to the same port it came from. 
Once the attacker learns if he is sharing the same physical server with his target, he can then mount additional targeted attacks~\cite{Ristenpart:2009:HYG:1653662.1653687}.

\subsection{Correlating Flow Rules and Network Policies}
\label{defence_inference}
\rew{Through the KYE attack, an attacker can} infer the exact network-level defense mechanism employed against specific attacks. 
In this section, we present a non-exhaustive set of defense policies~\cite{6419708} that are used in relation to our examples in Section~\ref{kye_sec}.
\rew{Furthermore, we explain how an attacker can correlate a sequence of flow rules obtained during the probing phase to the network policy they implement. }

\paragraph{Traffic Filtering}
One of the most basic network-level defense mechanisms is traffic filtering. 
A traffic filtering policy can be employed to mitigate a large range of attacks, including scanning and DoS attacks~\cite{Ambrosin:2015:LEM:2714576.2714612,entropy_based}. 
In SDN, traffic filtering is implemented simply by installing a \textit{drop} rule matching the offending network flows on OpenFlow switches. 
An attacker monitoring the flow table of an OpenFlow switch can easily detect the application of such defense mechanism. 
\rew{Indeed, before the policy is applied, the control plane will push on the OpenFlow switch normal flow rules, instructing the switch to forward the inbound traffic based on the destination address. }
When the attack is detected and the filtering is applied, the control plane will install only a single drop rule for the all the traffic coming from the attacker's IP address. 

\paragraph{Rate Limiting}
\label{rate_limiting_inference}
Rate limiting is a simple, yet effective defense mechanism to mitigate a wide variety of attacks like scanning and DoS~\cite{mehdi2011,Twycross:2003:ITV:1251353.1251373}. 
The most basic rate limiters, the ones assigning a maximum bandwidth to a given aggregate of network flows, are immediately recognizable for an attacker since they are directly defined in the flow rule matching the aggregate network flows~\cite{openflow}. 
More complex rate limiting approaches like those implemented in~\cite{mehdi2011,Twycross:2003:ITV:1251353.1251373} limit the rate of new network flows by delaying the installation of flow rules. 
\rw{In particular,~\cite{Twycross:2003:ITV:1251353.1251373} introduces the notion of working set:} for each given host, its working set is defined as the set of recently contacted hosts. 
Whenever a host creates a new network flow addressed at a host outside its working set, the controller will withhold the installation of the corresponding flow rule on the switch for a certain time. 
After this waiting time expires, the controller instructs the switch to forward the network flow without installing a flow rule. 
Only when the switch receives a positive reply from the destination host, the controller will install a new flow rule on the switch. 
An attacker can infer the presence of this defense mechanism by constantly probing the flow table of the switch after creating a new network flow. 
If rate limiting is in use, the attacker will notice that, even for extremely distant hosts, he will receive a response as soon as the flow rule is installed in the OpenFlow switch. 
Conversely, when no such technique is used, there will be a delay between the installation of the flow rule on the switch, which would happen as soon as the packet from the attacker reaches the OpenFlow switch, and the moment the reply is received. 
Therefore, by monitoring the delay in receiving a response after a flow rule is installed, the attacker can infer the use of this defense mechanism. 

\paragraph{Whitehole Network}
Another defense mechanism proposed in the literature is SYN proxy~\cite{Ambrosin:2015:LEM:2714576.2714612,Shin:2013:ASV:2508859.2516684}. 
\rw{SYN proxy techniques aim at countering the SDN-specific control plane saturation attack~\cite{Ambrosin:2015:LEM:2714576.2714612,Shin:2013:ASV:2508859.2516684}.
Additionally, SYN proxy implements a whitehole network~\cite{Shin:2013:ASV:2508859.2516684} at the switch level, providing mitigation also against network scans.} 
When such countermeasures are employed, the scenario is slightly different since these are pro-active techniques that are always active, rather than triggered by the installation of a flow rule. 
Even in this case though, the attacker can infer the existence and the exact type of the defense mechanism employed by the network. 
Indeed when SYN proxy techniques are used, the attacker will receive a response SYN-ACK packet \textit{without a flow rule being installed on the OpenFlow switch}~\cite{Shin:2013:ASV:2508859.2516684}. 
Additionally, the attacker will always receive a response SYN-ACK packet to each and every of his probes, even if directed to himself. 
This behaviour exposes the use of proxy techniques at the data plane level to the attacker, who can attack the OpenFlow switches through vulnerabilities of the proxy approach~\cite{Ambrosin:2015:LEM:2714576.2714612}.

\paragraph{Traffic Redirection}
Traffic redirection is a popular defense mechanism since it provides many opportunities to defend a system~\cite{Mahimkar:2007:DTN:1973430.1973454,6459946,FRESCO}. 
For instance, a defender can opportunistically route malicious traffic towards a honeypot, allowing it to isolate the attacker and to study his behaviour~\cite{FRESCO}. 
In SDN these defense mechanisms are easy to detect for an attacker. 
Indeed, by monitoring how the control plane updates flow rule entries for some given network flows, the attacker can infer if his attack traffic is diverted towards a security middlebox/honeypot, nullifying the effect of the countermeasure. 
In order to do so, an attacker first generates a new legit network flow towards a given destination $D_1$ which is routed through a port $P_i$ on the \switch. 
\rw{The attacker then repeats this step with different destinations, until for a given destination $D_n$ the controller pushes a flow rule instructing the switch to output the matching flow on a port $P_j <> P_i$.} 
As a second step, the attacker generates probing traffic with a high profile (e.g., high scanning or DoS rate) towards destinations $D_1$ and $D_n$ for a length of time, observing the flow rules installed in response. 
If traffic redirection techniques are in place, the control plane will install on the switch a new rule diverting the attack traffic towards the remote middlebox/honeypot. 
Therefore, all the attack traffic will be routed through the same output port on the OpenFlow switch. 
Since in the first phase the attacker selected the destinations $D_1$ and $D_n$ such that packets towards them would be outputted on different ports, if all attack packets towards those same destinations are tunneled through the same output port, then a redirection mechanism is present in the network.

\section{KYE Implementations}
\label{eval}
In order to prove the \rew{feasibility and effectiveness of the KYE attack, we implemented two instances of the attack on a test network.
In this section, we present the detailed implementation of our attacks, that were aimed at disclosing:}
\begin{enumerate}
        \item The presence of a scanning detection and defense mechanism in the network. If present, we also wanted to estimate the detection threshold. 
        \item The presence of a subnetwork access control mechanism. If present, we wanted to learn the subnetwork access control matrix.
\end{enumerate}

\begin{figure}[ht]
        \centering
	    \includegraphics[width=0.8\columnwidth]{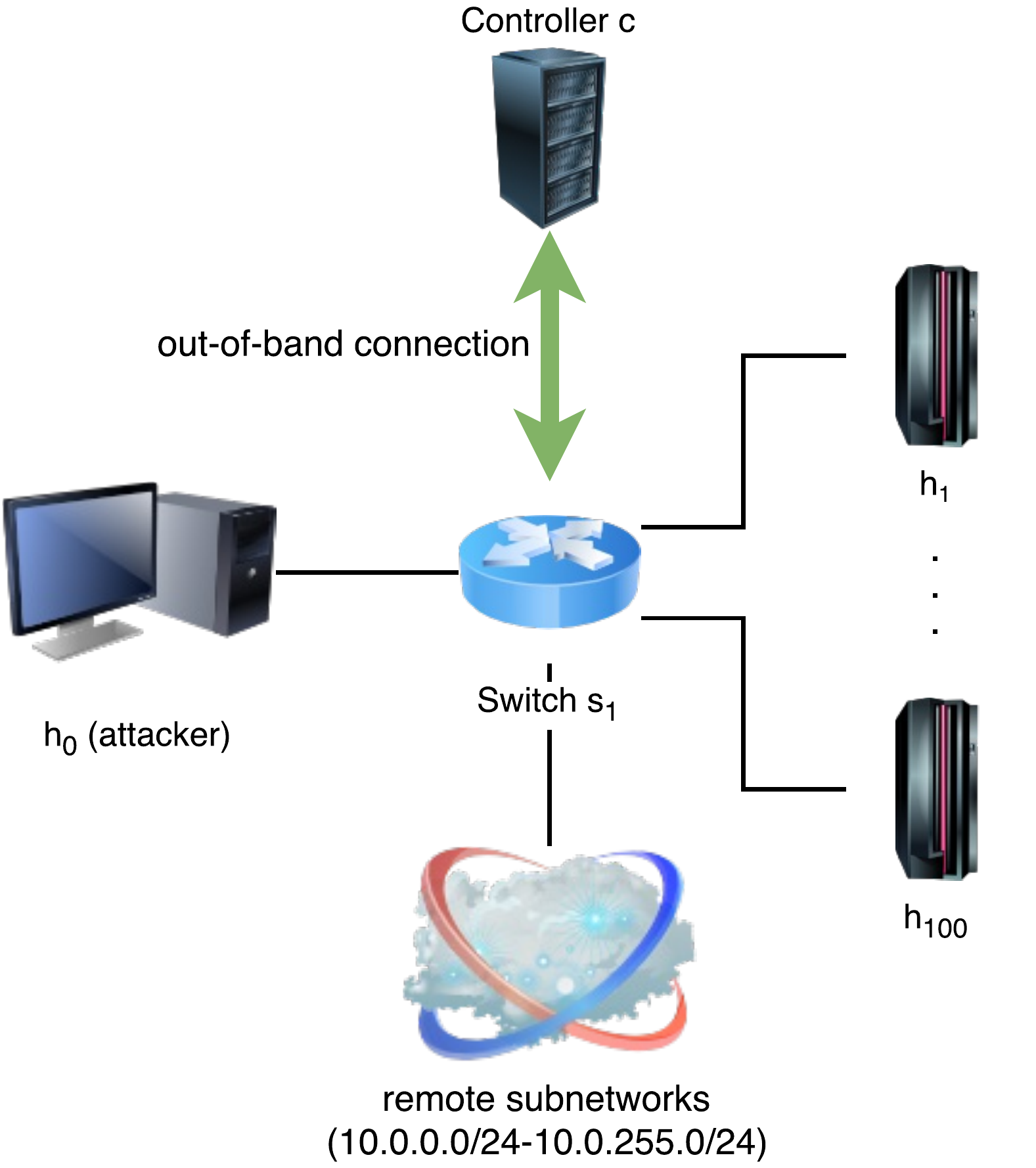}
	    \caption{Experimental setup used for the evaluation.}
	    \label{fig:experimental_setup}
	\end{figure}

Figure~\ref{fig:experimental_setup} depicts the setup used for the evaluation, which includes: the attacker $h_0$, a single OpenFlow switch $s_1$ connected to the controller $c$, and 100 legit hosts $h_1-h_{100}$. 
Hosts $h_1-h_{100}$ represent known web servers that always reply to connection requests. 
\rw{Hosts $h_1-h_{100}$, if needed, may be used by the attacker to obtain different connection success ratios during probing, and are not necessarily part of the target network.} 

In order to simplify the simulation, in our experiments the target network is comprised only of the switch $s_1$ and the network controller $c$. 
Even though in our test network we deployed only a single controller and a single switch, our experiments do not lose generality. 
This is because the logic used by the controller is the same that would be used in a more complex network, and the rules pushed by the controller are also the same. 
It is worth noting that it is not a requirement for the attacker to be directly connected to the OpenFlow switch, nor it is for hosts $h_1-h_{100}$. 
This is just a simplification we adopted in order to run our simulation. 
\rew{In this test network, we implemented the TRW-CB scanning detection algorithm, which is one of the most used anomaly detection algorithms~\cite{Ashfaq2008}. 
Our implementation of TRW-CB follows the SDN implementation detailed in~\cite{mehdi2011}. 
The test network also implements an IP-based access control mechanism.} 
We ran all our experiments in a simulated network using the Mininet network simulator~\cite{mininet} and the POX network controller~\cite{pox}. 

\subsection{\sloppy Disclosing Scanning Detection and Defense Mechanisms}
\label{exp_trw}
The target network implements TRW-CB at the control plane as a scanning detection mechanism, and traffic filtering as a defense measure (see Section~\ref{defence_inference}). 
TRW-CB employs both credit limiting, used to limit the amount of first contact connections pending, and sequential hypothesis testing to detect scanning hosts. 
In particular, we configured the TRW-CB algorithm with the same parameters used in the original paper~\cite{Schechter2004} (base credit of $10$, false positive rate~$\leq0.00005$, precision~$\geq0.99$). 

As an attacker, the first step is to learn if the target network has any kind of \rw{scanning} detection mechanism in place. 
To this end, we first initiated a scan with high packet/sec ratio using a spoofed IP, towards a remote subnet which is protected by the target SDN.
At the same time, we constantly monitored the flow table for a flow rule implementing a defense mechanism, which would indicate the presence of a \rw{scanning} detection mechanism. 
With our configuration, after the first 10 connection attempts failed, TRW-CB correctly identified our probes as scanning activity. 
Upon detection, the controller pushed a drop flow rule matching all packets coming from the attacker's IP address. 
Since we were monitoring the flow table of the switch, we detected the rule installation and concluded that the network indeed implements a detection mechanism for scanning attacks, as well as that it uses traffic filtering as a defense measure. 

\begin{figure}[tb]
	    \centering
	    \includegraphics[width=0.9\columnwidth]{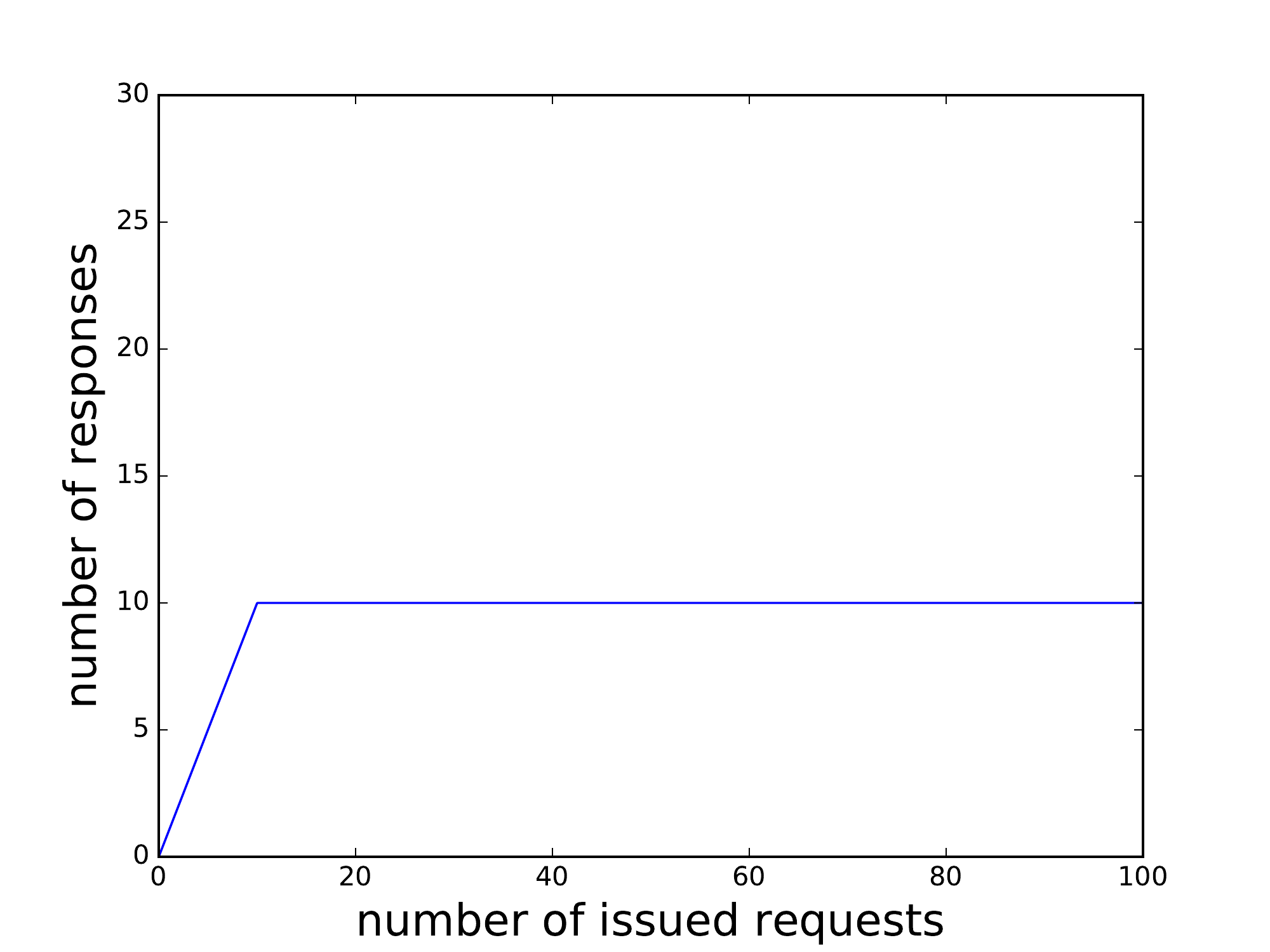}
	    \caption{Number of responses received vs. number of requests issued towards $h_1,h_{100}$ at each batch.}
	    \label{fig:credit_limit1}
	\end{figure}

After confirming the existence of a \rw{scanning} detection mechanism, the following step is to learn which detection criteria are used. 
As discussed in Section~\ref{scanning_inference}, in order to do so we initiate several batches of network scans with different characteristics, like scan rate and successful/failed connections ratio. 
The results of these batches of scans show two visible characteristics:

\begin{enumerate}
    \item The scanning activity is detected regardless of the scanning rate. This behaviour excludes rate-based scanning detection mechanisms. 
    \item For scan batches where connection requests were sent only towards $h_1,h_{100}$ (which should all send back a response), we received replies only from some of them, as illustrated in Figure~\ref{fig:credit_limit1}. 
\end{enumerate}

	\begin{figure}[tb]
	    \centering
	    \includegraphics[width=0.9\columnwidth]{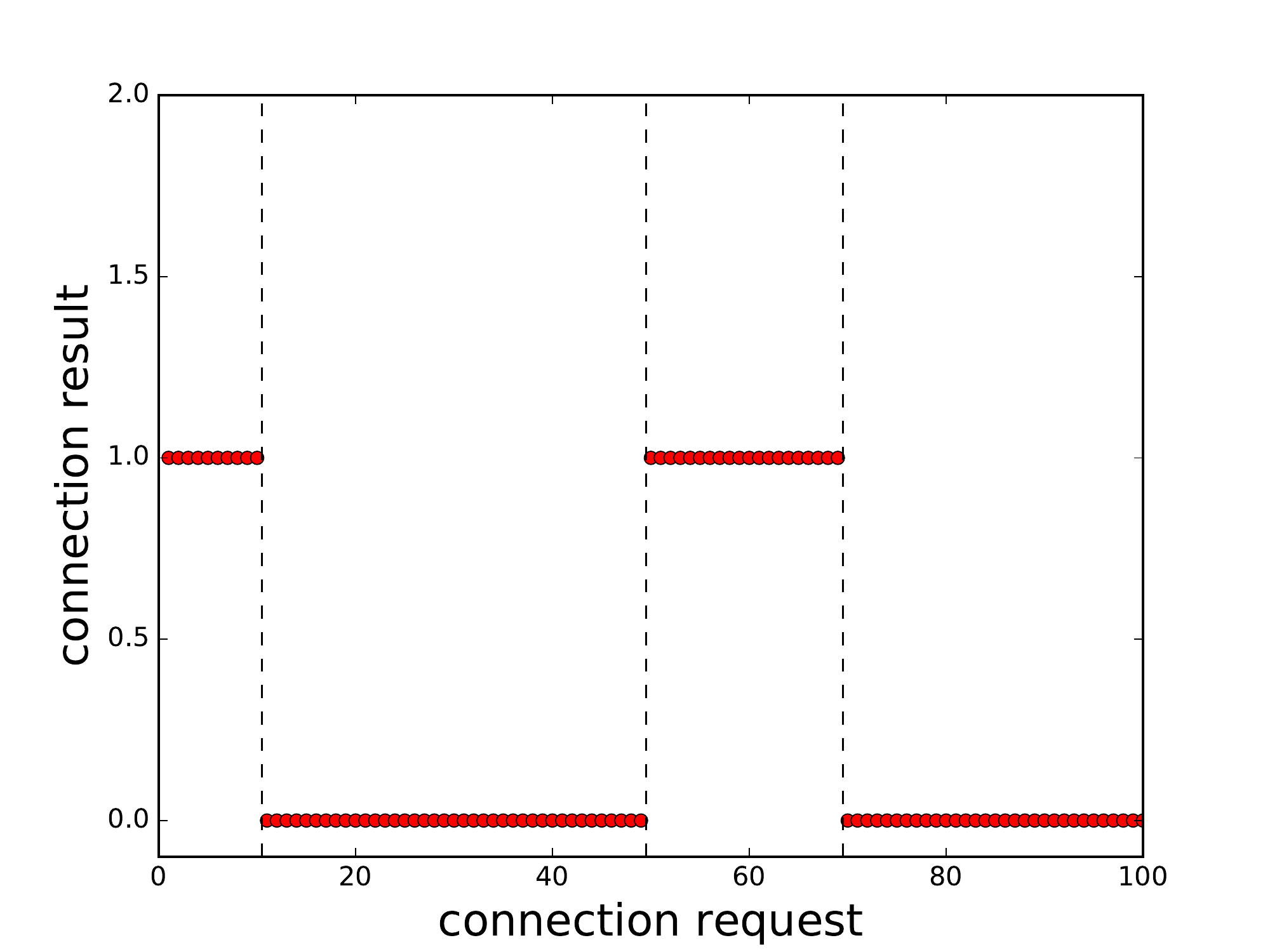}
        \caption{Connection result for each request of the batch sent towards a known host. 1 indicates a successful connection, 0 a failure (i.e., no response received).}
	    \label{fig:credit_limit2}
	\end{figure}

The behaviour shown in Figure~\ref{fig:credit_limit1} is consistent with rate limiting techniques, where connection requests sent at a rate above a certain threshold are dropped. 
To investigate this anomaly, we started a new scan towards $h_1,h_{100}$ with slightly lower rate. The results are illustrated in Figure~\ref{fig:credit_limit2}. 
As we can see, connections are allowed in bursts: replies were received for the first $10$ connections, then $39$ requests were dropped, after which connection attempts $50$ through $69$ were successful, and then connections were dropped once again. 
Since the scanning rate was constant for all the $100$ connection attempts, this behaviour excludes a standard rate limiting technique. 
Indeed, from Figure~\ref{fig:credit_limit2} we can see how a host is allowed to contact up to $10$ new hosts (i.e., $10$ starting credits), after which connections are blocked until pending replies are received. 
Upon receiving the replies, new credits are allocated to the host, whose connections are correctly forwarded once again.
This pattern is consistent with the presence of a credit based rate limiting mechanism~\cite{Schechter2004}. 

At this point, through the KYE attack, we learned that:
\begin{itemize}
    \item The network is using a detection mechanism for scanning, which is not rate based.
    \item The network is using drop rules as a defense mechanism against \rw{scanning}. 
        Moreover, the network employs an additional preemptive defense measure in the form of credit based rate limiting. 
        Each host is assigned a starting balance of $10$ credits and for each successful connection the hosts receives $2$ additional credits (the initial $10$ successful connections allowed $20$ more connections after replies were received). 
\end{itemize}

	\begin{figure}[tb!]
	    \centering
	    \includegraphics[width=0.9\columnwidth]{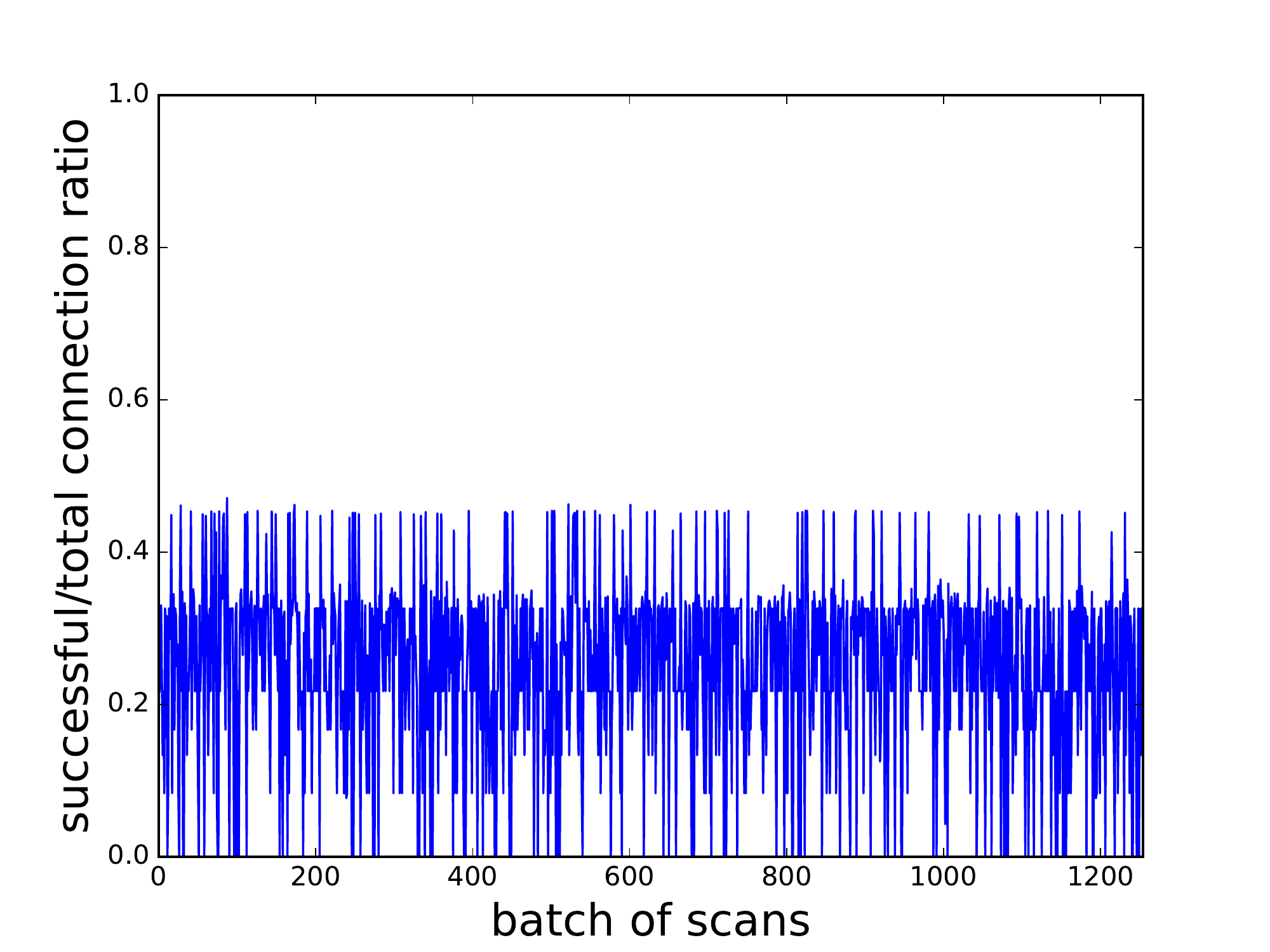}
        \caption{Ratio of the number of successful connections over the number of total connections, for each batch of scans which resulted in detection.}
	    \label{fig:connection_ratio}
	\end{figure}

	\begin{figure}[tb!]
	    \centering
	    \includegraphics[width=0.9\columnwidth]{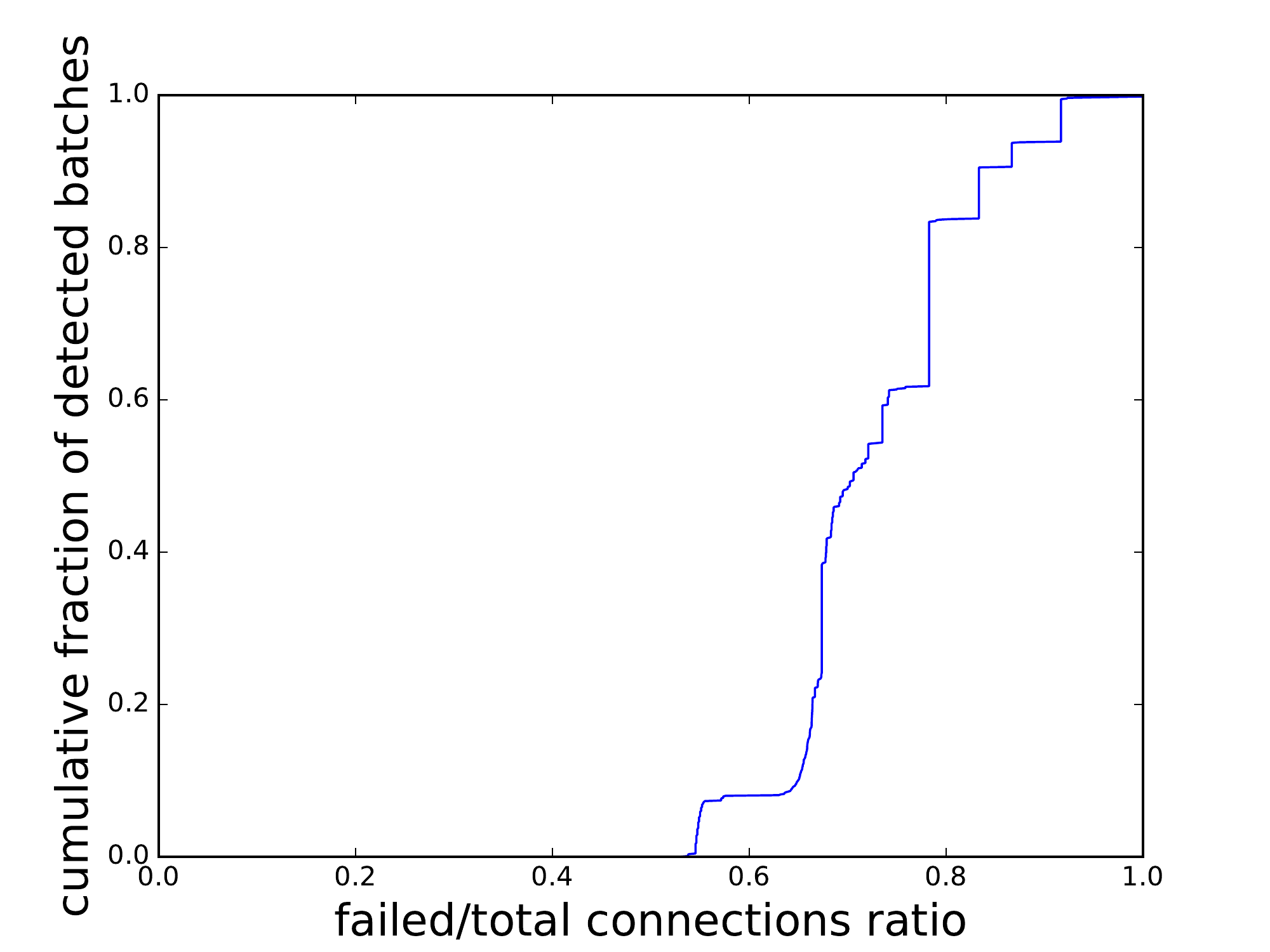}
        \caption{Empirical cumulative distribution function of the ratio of failed connections over total number of connections, for batches of scans which resulted in detection.}
	    \label{fig:cumulative_failed_ratio}
	\end{figure}

In the final step of the KYE attack, we initiated several batches of scans, with varying successful/failed connection ratios and scan duration. 
Each scanning batch terminated either after all planned scans were preformed, or abruptly upon detection. 
Since we are only interested in the characteristics of the scanning attacks that are detected, we isolate detected batches from undetected ones. 
\rw{For the batches of scans that were detected, figure~\ref{fig:connection_ratio} shows the ratio of successful connections over the total number of connections issued and Figure~\ref{fig:cumulative_failed_ratio} shows the cumulative distribution function of the ratio of failed connections over total number of connections.} 
As these two figures show, the scanning detection criteria employed by the network is clearly based on the ratio of successful and failed connections. 
Indeed, from Figure~\ref{fig:connection_ratio} we see that network scans are never detected when the ratio of successful connection over the total number of issued connections is above $\sim0.45$. 
Conversely, from Figure~\ref{fig:cumulative_failed_ratio} we can see that network scans are detected when the ratio of failed connections over total number of issued connections is above $\sim0.55$, and never detected when it is below that threshold.

\subsection{Disclosing the Subnetwork Access Control Matrix}
\label{exp_ac}
After learning the scanning detection criteria used by the network through a KYE attack, we show that we can also infer the complete access control matrix used by the network without being detected \rw{by the controller}. 
In our test network, we configured the POX controller with a set of static access control policies, where access to a certain subnetwork is allowed only from a subset of all subnetworks. 
Whenever a connection request from an unauthorized address is received, the controller instructs the switch to drop the packet without installing any flow rule. 
If the connection request is from an authorized address, the controller installs a normal forwarding flow rule on the switch for subsequent packets. 
In this setting, we perform a KYE attack, sending scan probes from each subnetwork to every other. 
\rew{Since with the previous attack we inferred the detection criteria used by the network for scans, \emph{we can now perform this network scan attack completely undetected}.}

The attack itself is very simple: at each scan probe, we spoof the source address to make it look like the source of the connection is part of a given subnetwork. 
We repeat the scan for each pair of source and destination remote subnetworks in the range $10.0.0.0\backslash24 - 10.0.255.0\backslash24$, while opening enough successful connections to remain below the detection threshold. 
After each scan, we read the flow table of the switch to detect which rule is installed. 
By monitoring the flow table, we see that no flow rules are installed when the source IP is not allowed to access the subnetwork, while a forwarding rule is installed when the access is authorized. 
Through this observation we are able to build the access control matrix illustrated in Table~\ref{tab:access_control_matrix}, which reflects exactly the access control rules that were set up at the network controller. 

    \begin{table*}[tb]
	    \small
    	\centering
    	\renewcommand{\tabcolsep}{5pt}
    	\begin{tabular}{|l|c|c|c|c|c|c|c|}
    		\hline
            &10.0.0.0/24 & 10.0.1.0/24 & 10.0.2.0/24 & 10.0.3.0/24 & 10.0.5.0/24 & 10.0.8.0/24 & 10.0.10.0/24 \\ \hline
	    	10.0.0.0/24		& \cmark & \cmark & \xmark & \xmark	& \xmark & \xmark & \xmark \\ \hline
    		10.0.1.0/24		& \cmark & \cmark & \xmark & \xmark	& \xmark & \xmark & \xmark \\ \hline
    		10.0.2.0/24		& \xmark & \xmark & \cmark & \cmark	& \xmark & \xmark & \xmark \\ \hline
	    	10.0.3.0/24		& \xmark & \xmark & \cmark & \cmark	& \xmark & \xmark & \cmark \\ \hline
	    	10.0.5.0/24		& \xmark & \xmark & \xmark & \xmark	& \cmark & \cmark & \xmark \\ \hline
	    	10.0.8.0/24		& \xmark & \xmark & \xmark & \xmark	& \cmark & \cmark & \xmark \\ \hline
	    	10.0.10.0/24	& \xmark & \xmark & \xmark & \cmark	& \xmark & \xmark & \cmark \\ \hline
    	\end{tabular}
    	\caption{Subnetwork access matrix for the target network, learned by the attacker through a KYE attack. Notation \cmark: access allowed; 
    	notation \xmark: access restricted.}
    	\label{tab:access_control_matrix}
    \end{table*}

\section{Countermeasure to the KYE Attack}
When considering potential countermeasures, the main problem is that the behaviour responsible for the vulnerability exploited by the KYE attack is also the main strength of SDN: programmability. 
Therefore, it is not possible to just remove or sensibly alter such behaviour without denaturating SDN itself. 
\rew{Additionally, classical techniques used to avoid eavesdropping, like randomized routing for instance~\cite{6682715}, are not applicable against the KYE attack (see Section~\ref{sec:rw}).} 

In this section, we propose a countermeasure to the KYE attack that does not require modifications to SDN, but rather that takes advantage of SDN programmability. 
We call this countermeasure \emph{flow obfuscation}. 
In this section, we use a stronger attack model than the one presented in Section~\ref{threat_model}. 
While in our presentation of the attack we used the threat model that is least favourable to the attacker (i.e., the attacker has a side-channel only for a single switch), for our countermeasure we consider a threat model that is least favourable to the defender: we assume that the attacker can obtain a flow table side-channel for up to $n$ switches in the target SDN network.

\subsection{Flow Obfuscation}
\begin{figure*}[t]
	    \centering
	    \includegraphics[width=0.75\textwidth]{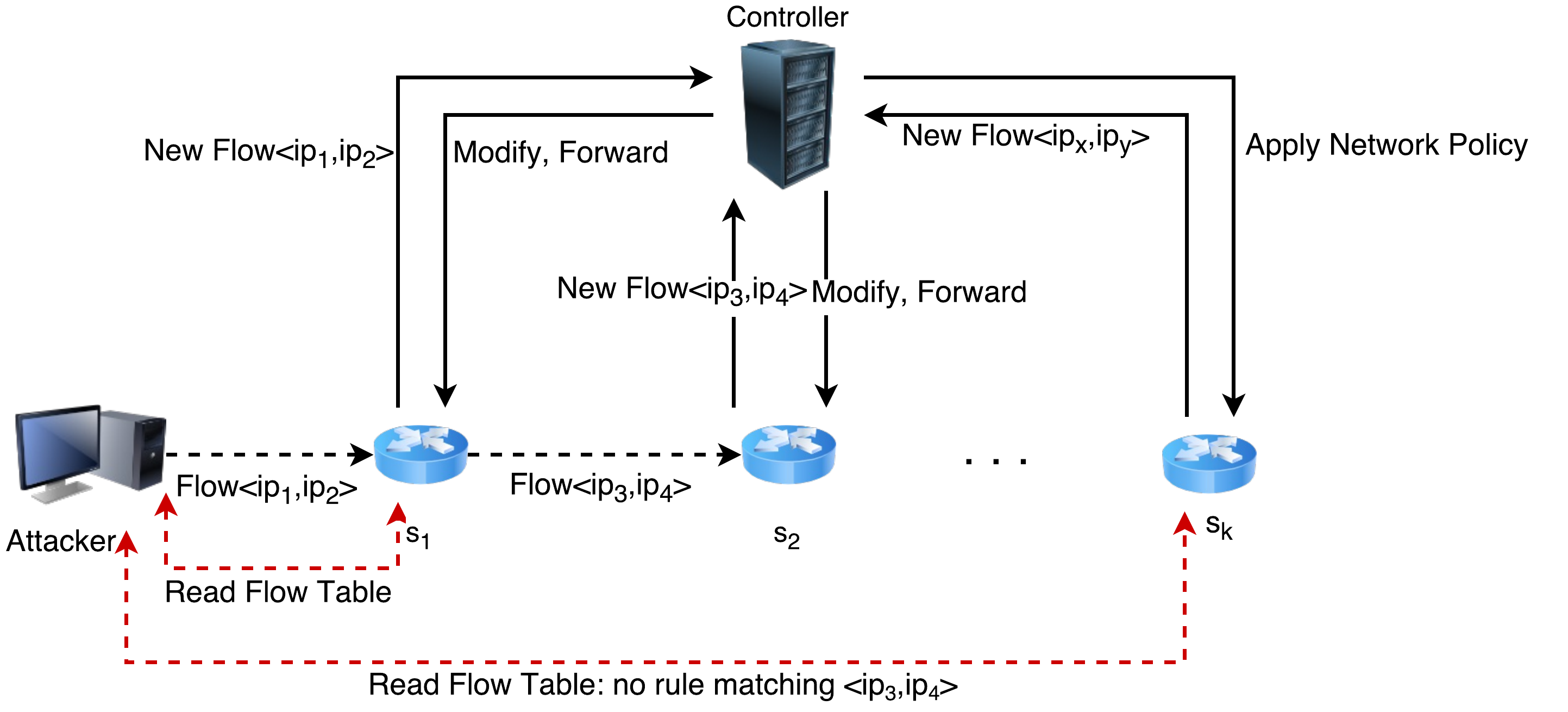}
        \caption{Overview of the flow obfuscation countermeasure. The red dashed lines indicate the flow table side-channel.}
	    \label{fig:flow_mod}
	\end{figure*}

In order to successfully mount a KYE attack, an attacker needs to be able to correlate (i) the network flows he generates to (ii) the reaction they cause in the network. 
In SDN, this corresponds to the installation of a specific flow rule that the attacker can detect, and from which he can obtain some knowledge about the network. 
Therefore, if it was possible to prevent the attacker from understanding which network flow caused the installation of which flow rule, the KYE attack would become unfeasible. 
In order to achieve this goal, we exploit the ability of OpenFlow switches to modify packets in transit. Figure~\ref{fig:flow_mod} provides an overview of this countermeasure. 

Any time a new network flow $f_i$ is received by a switch, the controller installs \emph{a single} flow rule on the switch, with two actions: 
the first action instructs the switch to modify some header fields of the packets of $f_i$ (e.g., source and destination IP), while the second action tells the switch which output port to use for packet forwarding. 
This process is repeated for the first $k-1$ switches $s_1,...,s_{k-1}$ in the path, after which, at switch $s_{k}$, the controller installs a rule to enforce the appropriate network policy for $f_i$. 
The goal of this countermeasure is to prevent the attacker from learning which network flow causes the installation of a given flow rule; 
indeed, since the attacker can control up to $n$ switches, when $k>n$ he will never be able to obtain the complete knowledge required for a successful KYE attack. 
In fact, when $k>n$ there are two possible scenarios:
\begin{itemize}
	\item If the attacker monitors $s_1,...,s_{k-1}$, then he does not monitor $s_{k}$, which is the switch applying the network policy flow rule.
	\item If the attacker monitors $s_{k}$, then he can not know if the installation of a rule on $s_k$ is caused by a network flow he generated.
        This is because the previous $k-1$ switches modify the packets at each step, and the attacker does not monitor all of them.
\end{itemize}

Please note that the use of source and destination IP to identify network flows is just used to present the countermeasure. 
In general, network flows can be matched on more or even completely different header fields. 
Even in such cases, the flow obfuscation countermeasure can be implemented by modifying these fields through \emph{set\_field} actions~\cite{openflow}.

\subsubsection{Choosing the Value of K}
While in order to prevent with $100\%$ probability all possible KYE attacks $k$ needs to be greater than $n$, in the general case this is not required. 
Indeed, depending on the average out-degree of the switches in the network, even with a value of $k\leq n$ the probability of the attacker obtaining a side-channel on the exact $k$ switches used for flow obfuscation can be low. 
Given an SDN network, let us assume that $o+1$ is the average out-degree of a switch and that an attacker knows the number of switches used in flow obfuscation. 
If an attacker $a$ can monitor at most $n$ switches (for simplicity, we assume $n\mod k=0$), the probability $P^s_{s1,sk}$ of him monitoring exactly the $k$ switches $s_1,...,s_{k}$ used for flow obfuscation is:

$$P^s_{s1,sk}=\biggl(\dfrac{(n/k)}o\biggr)^{k-1},$$

\noindent since attacker always knows the identity of $s_1$ (the edge switch through which the attacker's traffic is routed).
A network manager can decide the appropriate value for $k$ based on the expected value of $n$ and the maximum probability $P^s_{s1,sk}$ that he is willing to accept. 

\subsubsection{Limitations}
The flow obfuscation countermeasure can effectively prevent the KYE attack and is configurable to the needs of the network, but also presents some drawbacks. 
The first drawback is related to the delay in the application of network policies to network flows. 
Indeed, while in a normal SDN environment network policies are applied immediately at the ingress switch, when flow obfuscation is in place the network policy is delayed to the $k$-th hop in the network. 
While this behaviour might not impact all flows, in case of flows that require less than $k$ hops to reach the destination (e.g., flows that would be immediately dropped) it adds additional load on the network.  
The second drawback is related to the additional work required at the control plane. 
In fact, in order to correctly route the network flows the controller needs to keep track of each header modification applied by the switches. 
The higher $k$ is, the higher the amount of bookkeeping the controller has to do to correctly identify the network flows. 

These drawbacks can be mitigated by applying flow obfuscation in an intelligent manner. 
Indeed, the length of the flow obfuscation path can be tailored to the average length of network flows in the network; 
this way, on average flow obfuscation will add only a negligible delay to the application of network policies and the controller will incur a reduced overhead. 
Moreover, while on average flow obfuscation will indeed increase the load on network links, this load can be effectively distributed by the controller by dynamically changing the flow obfuscation path. 
This allows the controller to avoid link saturation, even in case of deliberate DoS attacks, thanks to the global network visibility provided by SDN. 
Additionally, it is worth noting that when applying flow obfuscation, the controller incurs overhead only for the first packet of each network flow. 
In fact, once the header-modification and forwarding flow rules are installed on the switches, all packets will be modified and correctly routed without the intervention of the controller.

\section{Limitations of the KYE Attack\\and Future Work}
The KYE attack requires constant assessments of the flow table by the attacker and the ability to recognise deviations in the type of flow rules installed. 
A human attacker can find this process bothersome and non-trivial in certain cases, leading to imprecise identification of detection conditions and/or defense measures applied. 
\rew{Additionally, the attacker might not be able to correctly identify the network policy applied if the policy itself is new or unknown to the community. 
While this might be the case, we argue that the attacker does not necessarily need to learn the exact nature of the network policy applied, as long as he is able to learn enough information for his purposes. 
For instance in case of network scanning, not being able to identify the defense mechanism used might be acceptable for the attacker, as long as he is able to understand when the scanning traffic is detected and when it is not. 
\rw{The attacker can learn when the scanning is detected} by observing a deviation in the flow rules installed for different types of probing traffic (e.g., very low scan rate probing traffic v.s. fast scan rate probing traffic). 
}

In our future work, we plan on automating the KYE attack by means of machine learning techniques. 
Indeed, the efficacy of the KYE attack is based on the ability to recognising patterns in the features of the generated traffic and in the flow rules installed in response, as well as to recognise significant deviations from such patterns. 
We believe that machine learning can be successfully applied to detect the baseline behaviour of the control plane under normal traffic conditions. 
Once this baseline behaviour has been identified, we can use a classifier to detect and categorize deviations from such behaviour triggered by specific traffic flows. 
It would then be possible to use another classifier to find the most relevant features of the attack traffic which caused the abnormal network state (e.g., detection of an attack). 
In this system model, the OpenFlow switch acts as an oracle: the switch can be repeatedly queried to learn if some network flows trigger a deviation in the usual type of flow rules installed.  
By analyzing the difference between the features of network flows that triggered the abnormal network state versus those which did not, it is possible to learn which are the characteristics that caused the reaction.

\rew{In complex real-world networks, it is possible to have several subnetworks, each with separate network policies. 
In this scenario, the KYE attack allows to learn the network configuration of the subnetwork for which the attacker has a flow table side-channel. 
If the attacker wishes to learn the configuration of other subnetworks, he will need to obtain a flow table side-channel for one switch on each of those subnetworks.
}

\section{Related Work}
\label{sec:rw}
SDN has become a popular research topic in recent years, especially in relation to security. 
Despite significant research efforts, to the best of our knowledge this is the first work to analyze the vulnerabilities caused by the distributed enforcement of rules in SDN. 
In~\cite{Kreutz:2013:TSD:2491185.2491199}, Kreutz et al. discuss the effects of an attacker compromising a switch in the network, which can result in traffic injection attacks, man-in-the-middle attacks and traffic filtering. 
\rew{Similarly, in~\cite{Antikainen2014} the authors consider an attacker with full control over the switch and discuss several possible attacks like man-in-the-middle, state, and topology spoofing. 
However, both of these works assume a stronger attacker, with full control over the switch, differently from us. 
Moreover, in these works~\cite{Antikainen2014,Kreutz:2013:TSD:2491185.2491199} the attacker actively modifies the state of the OpenFlow switch, exposing the attack to detection by the controller through querying~\cite{Benton:2013:OVA:2491185.2491222,Kamisinski:2015:FDM:2809826.2809833} the state of the switch or RTT analysis~\cite{Antikainen2014}.}
In~\cite{6733671}, Kl\"oti et al. analyze the overall security of OpenFlow using the STRIDE methodology. 
Between other threats, the authors also identify the risk of information disclosure in OpenFlow-based SDN. 
Indeed, they prove that by analyzing the RTT for a specific network flow, an attacker is able to infer if a flow rule for a specific network flow is already installed in an OpenFlow switch. 
\rw{While the goal of this attack is to obtain some information on the state of a switch, the KYE attack allows to learn a much greater amount of information about the logic and policies of the network, \rw{such as attack detection threshold and defense mechanisms applied.}} 
Another set of works related to ours pertain to SDN fingerprinting~\cite{Shin:2013:ASN:2491185.2491220}. 
SDN fingerprinting techniques use RTT to infer if a given network is an SDN or a classical network. 
These techniques apply the observation, made in~\cite{6733671}, that in SDN the first packet belonging to a network flow has a higher RTT then subsequent ones. 
By exploiting this asymmetry, an attacker can successfully infer if a network is an SDN with high accuracy. 
Fingerprinting attacks are related to our work in the sense that they also try to gather intelligence about the configuration of a network. 
\rew{However, the approach and the type of information obtained are completely different from our proposal and the amount of information retrieved with the KYE attack is much greater. 

The family of countermeasures that may seem applicable to the KYE attack are route randomization countermeasures~\cite{6682715}. 
These countermeasures aim to provide resistance to reconnaissance and eavesdropping. 
However, there are two major differences between the KYE attack and a classical eavesdropping scenario that make these countermeasures not applicable.
First, with the KYE attack, the  attacker only needs to have a flow table side-channel on the switch through which the attacker's traffic enters the network, rendering randomized routing ineffective. 
This is because, contrary to a classical eavesdropping scenario where the goal is to sniff the traffic of a victim, the attacker knows which switch in the network will be the gateway for his own traffic. 
Second, the duration of the network flow in case of the KYE attack and in case of classical eavesdropping are different. 
Indeed, while in a classical eavesdropping scenario the duration of the network flow is limited, in case of the KYE attack it can last as long as the attacker needs it to. 
Therefore even if the attacker's traffic is routed through the \switch only a fraction of the time, this simply means the attacker will have to increase the duration of the probing phase of the KYE attack. 
In particular, the attacker can opportunistically suspend the probing when the traffic is not routed through the \switch anymore, and resume when the traffic is routed through it once again. 
Finally, the KYE attack can be used to learn the characteristics of the randomized routing mechanism itself, like the duration of the randomization intervals.}

\section{Conclusions}
In this paper, we proposed a thorough analysis of the vulnerability introduced by the on-demand installation of flow rules in SDN. 
We presented the novel KYE attack which, with minimal requirements, allows an adversary to gather an extensive amount of information regarding the configuration of the network, ranging from security-related aspects to network engineering policies. 
We \rw{implemented the KYE attack and conducted a} thorough evaluation, showing its feasibility against a popular scanning detection algorithm and against standard access control policies. 
Finally, we proposed the flow obfuscation countermeasure to the KYE attack, which provides provable security guarantees and can be tailored to the needs of a specific network under consideration.

\bibliographystyle{abbrv}
\bibliography{bk_bibliography}  %
\end{document}